\documentclass[twocolumn,showpacs,preprintnumbers,amsmath,amssymb,superscriptaddress,pre]{revtex4}
\usepackage{graphicx}% Include figure files
\usepackage{dcolumn}% Align table columns on decimal point
\usepackage{bm}% bold math

\newcommand{\unit}[1]{\,\mathrm{#1}}
\newcommand{\e}{\mathrm{e}}
\newcommand{\ep}{\varepsilon}
\newcommand{\expct}[1]{\langle #1 \rangle}
\newcommand{\pr}{{\bm{r}}}

\begin{document}

%\preprint{APS/123-QED}

\title{Experimental realization of directed percolation criticality in turbulent liquid crystals}

\author{Kazumasa A. Takeuchi}
 \email{kazumasa@daisy.phys.s.u-tokyo.ac.jp}
\affiliation{
Department of Physics, The University of Tokyo, 7-3-1
  Hongo, Bunkyo-ku, Tokyo, 113-0033, Japan
}%
\affiliation{
Service de Physique de l'\'Etat Condens\'e, CEA-Saclay, 91191 Gif-sur-Yvette, France
}%
\author{Masafumi Kuroda}
\affiliation{
Department of Physics, The University of Tokyo, 7-3-1
  Hongo, Bunkyo-ku, Tokyo, 113-0033, Japan
}%

\author{Hugues Chat\'e}
\affiliation{
Service de Physique de l'\'Etat Condens\'e, CEA-Saclay, 91191 Gif-sur-Yvette, France
}%

\author{Masaki Sano}
 \email{sano@phys.s.u-tokyo.ac.jp}
\affiliation{
Department of Physics, The University of Tokyo, 7-3-1
  Hongo, Bunkyo-ku, Tokyo, 113-0033, Japan
}%

\date{\today}% It is always \today, today,

\begin{abstract}
This is a comprehensive report on the phase transition
 between two turbulent states
 of electroconvection in nematic liquid crystals,
 which was recently found by the authors
 to be in the directed percolation (DP) universality class
 [K. A. Takeuchi \textit{et al.}, Phys. Rev. Lett. \textbf{99}, 234503 (2007)].
We further investigate both static and dynamic critical behavior
 of this phase transition,
 measuring a total of 12 critical exponents, 5 scaling functions,
 and 8 scaling relations,
 all in full agreement with those characterizing the DP class
 in 2+1 dimensions.
Developing an experimental technique to create
 a seed of topological-defect turbulence by pulse laser,
 we confirm in particular the rapidity symmetry,
 which is a basic but nontrivial consequence
 of the field-theoretic approach to DP.
This provides the first clear experimental realization of this outstanding,
 truly out-of-equilibrium universality class, dominating most phase transitions
 into an absorbing state.

\end{abstract}

\maketitle

\section{introduction}  \label{sec:Introduction}

Absorbing states,
 i.e., states which systems may fall into but never escape from,
 and phase transitions into them are expected to be ubiquitous in nature.
For instance, spreading or contamination processes
 like fires or epidemics exhibit such transitions
 when the propagation rate changes:
 initially active (infected) regions eventually disappear forever at low rates,
 i.e., the absorbing state is reached,
 while they can be sustained (pandemic regime) for fast-enough propagation.
Examples abound far beyond:
 hundreds of numerical models,
 describing, e.g., catalytic reactions, granular flows,
 and calcium dynamics in living cells, to name but a few,
 have been shown to exhibit such absorbing phase transitions
 \cite{Hinrichsen-AdvPhys2000, Henkel_etal-Book2008}.
Such phase transitions also naturally arise from general problems such as
 synchronization \cite{Baroni_etal-PRE2001, Ahlers_Pikovsky-PRL2002},
 self-organized criticality \cite{Dickman_etal-BrazJPhys2000},
 spatiotemporal intermittency \cite{Pomeau-PhysD1986},
 and depinning \cite{Buldyrev_etal-PRA1992,Ginelli_etal-PRE2003}.
 The vast majority of these transitions
% are continuous and
 share the same critical behavior,
 that of the ``directed percolation'' (DP) class
 \cite{Hinrichsen-AdvPhys2000, Henkel_etal-Book2008},
 as long as they are continuous.

Deep theoretical issues underpin this situation.
Whereas universality is well understood for systems in 
 thermodynamic equilibrium, this is still not the case for systems driven
 out of equilibrium,
 where even the relevant ingredients determining the class
 are not well understood. 
In this context, absorbing phase transitions are central because 
 of their genuine nonequilibrium character,
 since absorbing states directly imply violation of the detailed balance.
Janssen and Grassberger \cite{Janssen-ZPhysB1981, Grassberger-ZPhysB1982}
 conjectured that
 the DP universality class 
 contains all continuous transitions into a single effective absorbing state
 in the absence of any extra symmetry or conservation law. The DP class
 thus appears as the simplest, and most common case, as testified by 
 overwhelming numerical evidence.

However, the situation has been quite different in experiments.
Over the last twenty years and more,
 a number of experiments have been performed
% on absorbing phase transitions
% that are theoretically expected to be in the DP class,
 in situations where DP-class transitions would be theoretically expected,
 but they have always yielded mixed and/or partial results
 with limited accuracy
 \cite{Ciliberto_Bigazzi-PRL1988,Daviaud_etal-PRA1990,Buldyrev_etal-PRA1992,Michalland_etal-EurophysLett1993,Willaime_etal-PRE1993,Degen_etal-PRE1996,Colovas_Andereck-PRE1997,Tephany_etal-PhysA1997,Daerr_Douady-Nature1999,Cros_LeGal-PhysFluids2002,Rupp_etal-PRE2003,Lepiller_etal-PhysFluids2007,Pirat_etal-PRL2005} (Table \ref{tbl:EarlierExperiments}).
This lack of fully convincing experimental realizations,
 in contrast with the wealth of numerical results, has been found surprising
and a matter of concern  in the literature
 \cite{Grassberger-Monograph1997,Hinrichsen-AdvPhys2000}.

Recently, though, studying a spatiotemporal intermittency regime occurring
in turbulent liquid crystals, we found a transition whose complete
set of static critical exponents match those of the DP class \cite{Takeuchi_etal-PRL2007}.
The goal of the present paper is to provide a comprehensive report on this transition
including not only a more complete description of experiments designed to 
investigate the static critical behavior, but also new experiments giving access to 
dynamic critical behavior.

\begin{table*}
 \begin{minipage}{\textwidth}
  \caption{Summary of critical exponents measured in earlier experiments.\protect\footnotemark[1]}
  \label{tbl:EarlierExperiments}
  \catcode`?=\active \def?{\phantom{0}}
  \begin{tabular}{lllllllll} \hline \hline
   (1+1)D System & Size\footnotemark[2]~~ & $\beta$ & $\nu_\perp$ & $\nu_\parallel$ & $\mu_\perp$ & $\mu_\parallel$ & Refs. \\ \hline
   annular Rayleigh-B\'enard & $?22$ & & $0.5$ & & $1.9(1)$ & $1.9$ &\cite{Ciliberto_Bigazzi-PRL1988} \\
   annular Rayleigh-B\'enard & $?35$ & & $0.5$ & $0.5$ & $1.7(1)$ & $2.0(1)$ &\cite{Daviaud_etal-PRA1990} \\
   linear Rayleigh-B\'enard & $?25.7$ & $0.30(5)$ & $0.50(5)$ & $0.50(5)$ & $1.6(2)$ & $2.0(2)$ &\cite{Daviaud_etal-PRA1990} \\
   interface roughening & & & \multicolumn{2}{l}{$\nu_\perp/\nu_\parallel = 0.63(4)$} & & &\cite{Buldyrev_etal-PRA1992} \\
   viscous fingering & $?60$ & $0.45(5)$ & & & $0.64(2)$ & $0.61(2)$ &\cite{Michalland_etal-EurophysLett1993} \\
   vortices of fluid & $?15$ & $0.5$ & & & & $1.7$ &\cite{Willaime_etal-PRE1993} \\
   Taylor-Deen & $?90$ & $1.30(26)$ & $0.64, 0.53$ & $0.73$ & $1.67(14)$ & $1.74(16)$ &\cite{Degen_etal-PRE1996} \\
   Taylor-Couette & $?70$ & $1$ & $0.4$ & & $1.4$-$2.5$ & &\cite{Colovas_Andereck-PRE1997} \\
   granular flow & & & \multicolumn{2}{l}{$\nu_\parallel-\nu_\perp = 1$} & & & \cite{Daerr_Douady-Nature1999,Hinrichsen_etal-PRL1999}\\
   torsional Couette & & $0.30(1)$ & $0.53(5)$ & & & & \cite{Cros_LeGal-PhysFluids2002} \\
   ferrofluidic spikes & $108$ & $0.30(5)$ & $1.1(2)$ & $0.62(14)$ & $1.70(5)$ & $2.1(1)$ & \cite{Rupp_etal-PRE2003} \\
   lateral heat convection in annulus & $114$ & $0.27(3)$ & $0.30(4)$ & $0.75(3)$ & & & \cite{Lepiller_etal-PhysFluids2007} \\ \hline
   DP\footnotemark[3] & & $0.276$ & $1.097$ & $1.734$ & $1.748$\footnotemark[4] & $1.841$\footnotemark[4] & \cite{Jensen-JPhysA1999}\\
   & & \multicolumn{2}{l}{$\nu_\perp/\nu_\parallel = 0.633,$} & \multicolumn{2}{l}{$\nu_\parallel-\nu_\perp = 0.637$} & & \\ \hline\hline
   (2+1)D System & Size & $\beta$ & $\nu_\perp$ & $\nu_\parallel$ & $\mu_\perp$ & $\mu_\parallel$ & Refs. \\ \hline
   liquid columns & $169$ & $0.56(5)$ & & & & &\cite{Pirat_etal-PRL2005} \\
   DSM1-DSM2 (present paper)\footnotemark[5] & $2.7 \times 10^6$ & $0.59(4)$ & $0.75(6)$ & $1.29(11)$ & $1.08(18)$ & $1.60(5)$ & ----- \\ \hline
   DP\footnotemark[3] & & $0.583(3)$ & $0.733(3)$ & $1.295(6)$ & $1.204(2)$\footnotemark[4] & $1.5495(10)$\footnotemark[4] & \cite{Grassberger_Zhang-PhysA1996,Voigt_Ziff-PRE1997} \\ \hline\hline
  \footnotetext[1]{Number in parentheses is the range of error given by the corresponding authors.}
  \footnotetext[2]{Number of effective degrees of freedom indicated by the corresponding authors.}
  %\footnotetext[3]{These values are estimated using scaling relations $\mu_{\perp,\parallel} = 2 - \beta/\nu_{\perp,\parallel}$, which might be slightly violated (in the order of $10^{-2}$) due to intermittency in DP \cite{Hede_etal-JSP1991, Henkel_Peschanski-NuclPhysB1993}.}
  \footnotetext[3]{Some exponents are obtained using scaling relations. Errors are then estimated by the law of propagation of error.}
  \footnotetext[4]{See also the remark \cite{Note4}.}
  \footnotetext[5]{Only some of the measured exponents are shown. See Table \ref{tbl:SummaryExponents} for the complete list.}
  \end{tabular}
 \end{minipage}
\end{table*}

The paper is organized as follows.
We first illustrate the coarse-grained dynamics
 of the turbulent regime of electroconvection studied
 and our basic experimental setup (Sec.\ \ref{sec:Setup}),
 with further details on image analysis given in the Appendix.
To characterize the critical behavior,
 we perform three series of experiments:
 (a) steady-state experiment under constant applied voltages
 (Sec.\ \ref{sec:SteadyState}),
 (b) critical-quench experiment starting from fully active initial conditions
 (Sec.\ \ref{sec:CriticalQuench}),
 and (c) critical-spreading experiment starting from a single active seed,
 prepared with a novel experimental technique using pulse laser,
 developed in this work
 (Sec.\ \ref{sec:CriticalSpreading}).
In Sec.\ \ref{sec:Summary},
 our results are summarized and we discuss why
 clear DP-class critical behavior is observed rather easily in our system, 
 contrary to many other experiments performed in the past in this context.

\section{experimental setup}  \label{sec:Setup}

\begin{figure*}[t]
 \includegraphics[clip]{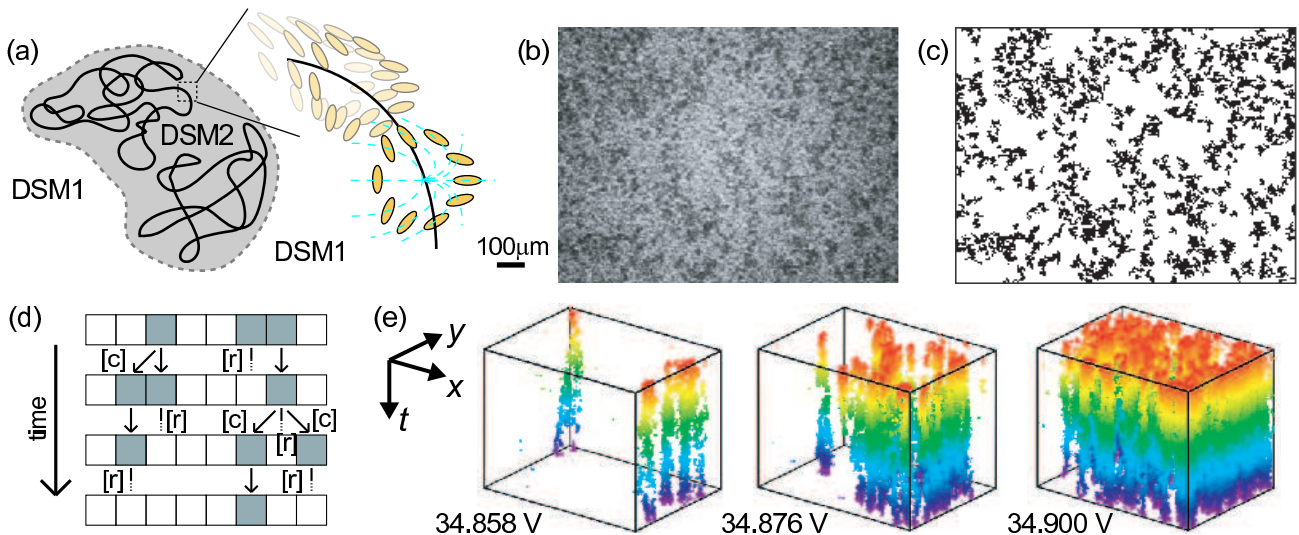}
 \caption{(Color online) Spatiotemporal intermittency between DSM1 and DSM2. (a) Sketch of a DSM2 domain with many entangled disclinations, i.e., loops of singularities in orientations of liquid crystal. Blue dashed curves in the close-up indicate contour lines of equal alignment. (b) Snapshot taken at $35.153\unit{V}$. Active (DSM2) patches appear darker than the absorbing DSM1 background. See also Movie S1 in Ref.\ \cite{EPAPS}. (c) Binarized image of (b). See also Movie S2 in Ref.\ \cite{EPAPS}. (d) Sketch of the dynamics: DSM2 domains (gray) stochastically contaminate [c] neighboring DSM1 regions (white) and/or relax [r] into the DSM1 state, but do not nucleate spontaneously within DSM1 regions (DSM1 is absorbing). (e) Spatiotemporal binarized diagrams showing DSM2 regions for three voltages near the critical point. The diagrams are shown in the range of $1206\unit{\mu{}m} \times 899\unit{\mu{}m}$ (the whole observation area) in space and $6.6\unit{s}$ in time.}
 \label{fig:DSM1DSM2}
\end{figure*}

We work on the electrohydrodynamic convection of nematic liquid crystals,
 which occurs when a thin layer of liquid crystal
 is subjected to an external voltage strong enough to trigger
 the Carr-Helfrich instability
 \cite{Dubois-Violette_etal-JPhysParis1971, deGennes_Prost-Book1993}.
This is \textit{a priori} a suitable system to study critical behavior
 thanks to its possibly large aspect ratio, fast response time,
 and easy controllability.
We focus on the transition between two turbulent regimes,
 called dynamic scattering modes 1 and 2 (DSM1 and DSM2),
 observed successively upon increasing
 the root-mean-square amplitude of the voltage $V$
 at relatively low frequencies
 \cite{deGennes_Prost-Book1993,Kai_etal-JPSJ_PRL}.
The difference between DSM1 and DSM2 lies
 in their density of topological defects in the director field
 [Fig.\ \ref{fig:DSM1DSM2}(a)].
In the DSM2 state, a large quantity of these defects,
 called disclinations, are present \cite{Nehring-PRA1973}.
They elongate and split constantly
 under the shear due to the fluctuating turbulent flow around.
In DSM1, on the other hand,
 disclinations are present but kept smaller than the critical size
 and disappear immediately.
Their density thus remains very low.
The many disclinations in DSM2
 lead to the loss of macroscopic nematic anisotropy
 and to a lower light transmittance than in DSM1.

\begin{figure*}[t]
 \includegraphics[clip]{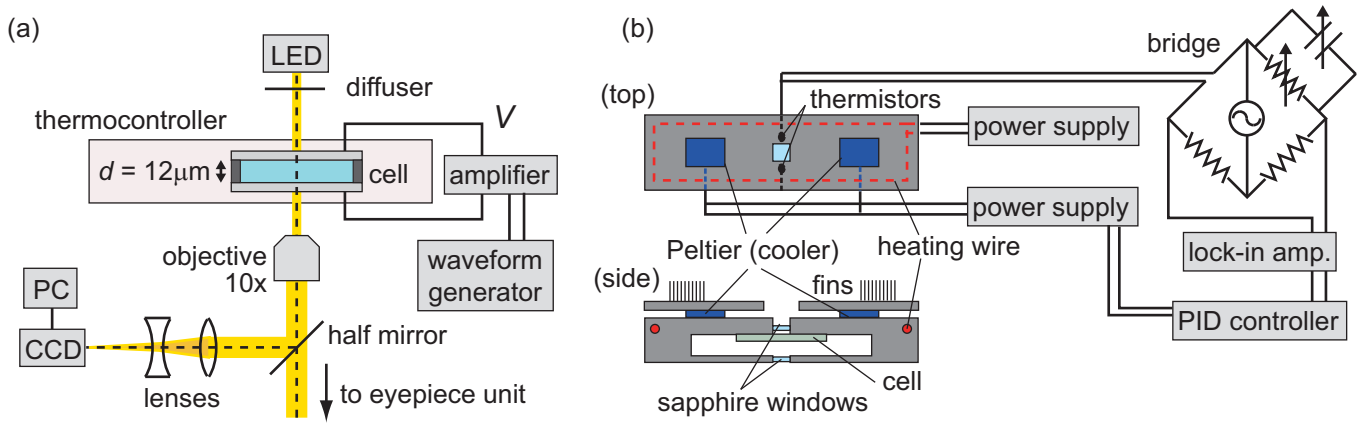}
 \caption{(Color online) Schematic diagram of the experimental setup, in its entirety (a) and for the thermocontroller (b). LED: light-emitting diode, CCD: charge-coupled device camera, PC: computer, PID: proportional-integral-derivative. See text for details.}
 \label{fig:Setup}
\end{figure*}

Our basic experimental setup
% which is a standard one for electroconvection,
 is shown in Fig.\ \ref{fig:Setup}.
The sample cell is made of two parallel glass plates
 spaced by a polyester film of thickness $d=12\unit{\mu{}m}$.
Both inner surfaces are covered
 with
% indium-tin oxide
 transparent electrodes
 of size $14\unit{mm} \times 14\unit{mm}$,
 coated with polyvinyl alcohol and then rubbed
 in order that molecules are planarly aligned in the $x$ direction,
 defined thereby.
The cell is filled with $N$-(4-methoxybenzylidene)-4-butylaniline
 (MBBA; purity $> 99.5\%$, Tokyo Chemical Industry)
% with purity greater than 99.5\%,
 doped with 0.01 wt.\% of tetra-$n$-butylammonium bromide.
The temperature of the cell is kept constant
 carefully by a handmade thermocontroller,
 composed of heating wires and Peltier elements
 controlled by a proportional-integral-derivative
 feedback loop with a lock-in amplifier
% (Model 5302, EG \& G Princeton Applied Research)
 [Fig.\ \ref{fig:Setup}(b)].
Windows of the thermocontroller are made of sapphire
 in order to improve the spatial homogeneity of the temperature.
Throughout each series of experiments,
 the cell temperature is maintained at $26.0\unit{^\circ{}C}$
 with fluctuations typically of a few $\unit{mK}$,
 unless otherwise stipulated,
 measured by three thermistors placed at different positions close to the cell.

We observe the electroconvection through the transmitted light
 from a handmade stabilized light source made of light-emitting diodes,
 recorded by a charge-coupled device camera.
The observed region is a central rectangle of size
 $1217 \unit{\mu{}m} \times 911 \unit{\mu{}m}$ (Fig.\ \ref{fig:Setup}, inset).
% Given the minimum linear size of DSM2 domains,
Since there is a minimum linear size of DSM2 domains,
 $d/\sqrt{2}$
 \cite{Kai_etal-JPSJ_PRL},
 we can roughly estimate the number of effective degrees of freedom
 at $1650 \times 1650 \approx 2.7 \times 10^6$ for the convection area
 and $143 \times 107 \approx 1.5 \times 10^4$ for the observation area. 
Note that the meaningful figure is that of the total
 system size, which is at least four orders of magnitude larger
 than in earlier experimental studies (Table \ref{tbl:EarlierExperiments}).
In the following, we vary $V$ and fix the frequency at $250\unit{Hz}$,
at roughly one third of the cutoff frequency $820 \pm 70\unit{Hz}$
which separates the conducting and the dielectric regimes of electroconvection
 \cite{Dubois-Violette_etal-JPhysParis1971, deGennes_Prost-Book1993}.

Disclinations being topological defects,
the spontaneous nucleation of DSM2 in a DSM1 domain is in principle forbidden.
It is indeed an essentially unobservable rare event,
 except along the edges of the electrode and for very high voltages,
 far from the range investigated in the present paper.
Therefore, the fully DSM1 state serves as an absorbing state.
On the other hand, 
 DSM2 domains introduced externally, or present initially in the
system can remain sustained in the bulk for large enough voltages,
 but eventually disappear for voltages lower
 than a certain threshold $V_{\rm c}$.
Closely above  $V_{\rm c}$, a regime of spatiotemporal intermittency (STI) is observed,
 with DSM2 patches moving around on a DSM1 background
 [Fig.\ \ref{fig:DSM1DSM2}(b), and Movie S1 of Ref.\ \cite{EPAPS}].
The basic dynamics of the observed STI is illustrated
 in Fig.\ \ref{fig:DSM1DSM2}(d):
 active DSM2 patches evolve in space-time essentially by contamination
 of neighboring inactive (absorbing) DSM1 regions
 and by relaxation into the DSM1 state.
This suggests an absorbing phase transition
 induced by change in rates of both elementary processes
 \cite{Pomeau-PhysD1986},
 which are functions of the applied voltage here.
The order parameter $\rho$ is then simply the ratio of the surface
 occupied by active DSM2 regions to the whole area.

Prior to any analysis,
 we must distinguish DSM2 domains from DSM1.
This binary reduction can be easily performed by our eyes,
 so we automated it, based on the facts that
 DSM2 domains have lower transmittance,
 longer time correlations,
 and have a minimum area of $d^2/2$ \cite{Kai_etal-JPSJ_PRL}
 (see Appendix for details).
A typical result is shown in Fig.\ \ref{fig:DSM1DSM2}(c),
 and in Movie S2 of Ref.\ \cite{EPAPS}.
Figure \ref{fig:DSM1DSM2}(e) displays spatiotemporal diagrams
 obtained this way, showing how DSM2 patches evolve
 in the steady state.
This supports the qualitative dynamics illustrated
 in Fig.\ \ref{fig:DSM1DSM2}(d) and indeed looks
 like the directed percolation of, say, water in a porous medium 
under gravitational field.

\section{steady-state experiment}  \label{sec:SteadyState}

We first observe STI in the steady state under constant voltage $V$,
 in the range of $34.858\unit{V} \leq V \leq 39.998\unit{V}$.
The voltage for the onset of steady roll convection (Williams domain)
 is $V^* = 8.95\unit{V}$.
Spatiotemporal distributions of DSM2 patches are recorded
 over the period $1000\unit{s} < T < 8000\unit{s}$,
 which is longer than $10^3$ correlation times
 defined from the fluctuations of the order parameter $\rho(t)$.

\begin{figure}[t]
 \includegraphics[clip]{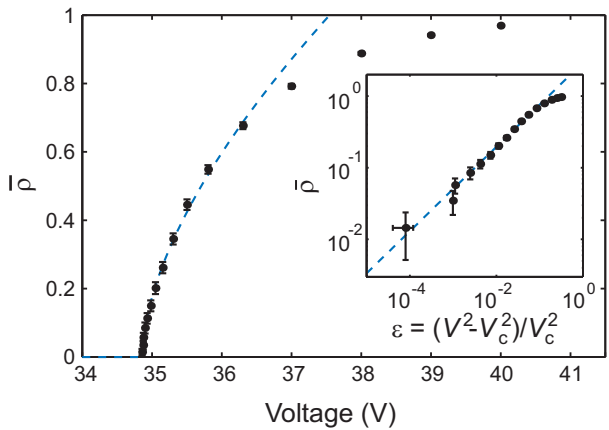}
 \caption{(Color online) The average DSM2 fraction $\bar\rho$ as a function of $V$ in the steady state. Inset: same data in logarithmic scales. Errorbars indicate the standard deviation of fluctuations in $\rho(t)$ and $V$. Blue dashed lines are fitting curves.}
 \label{fig:SteadyStateRho}
\end{figure}%

\begin{figure*}[t]
 \includegraphics[clip]{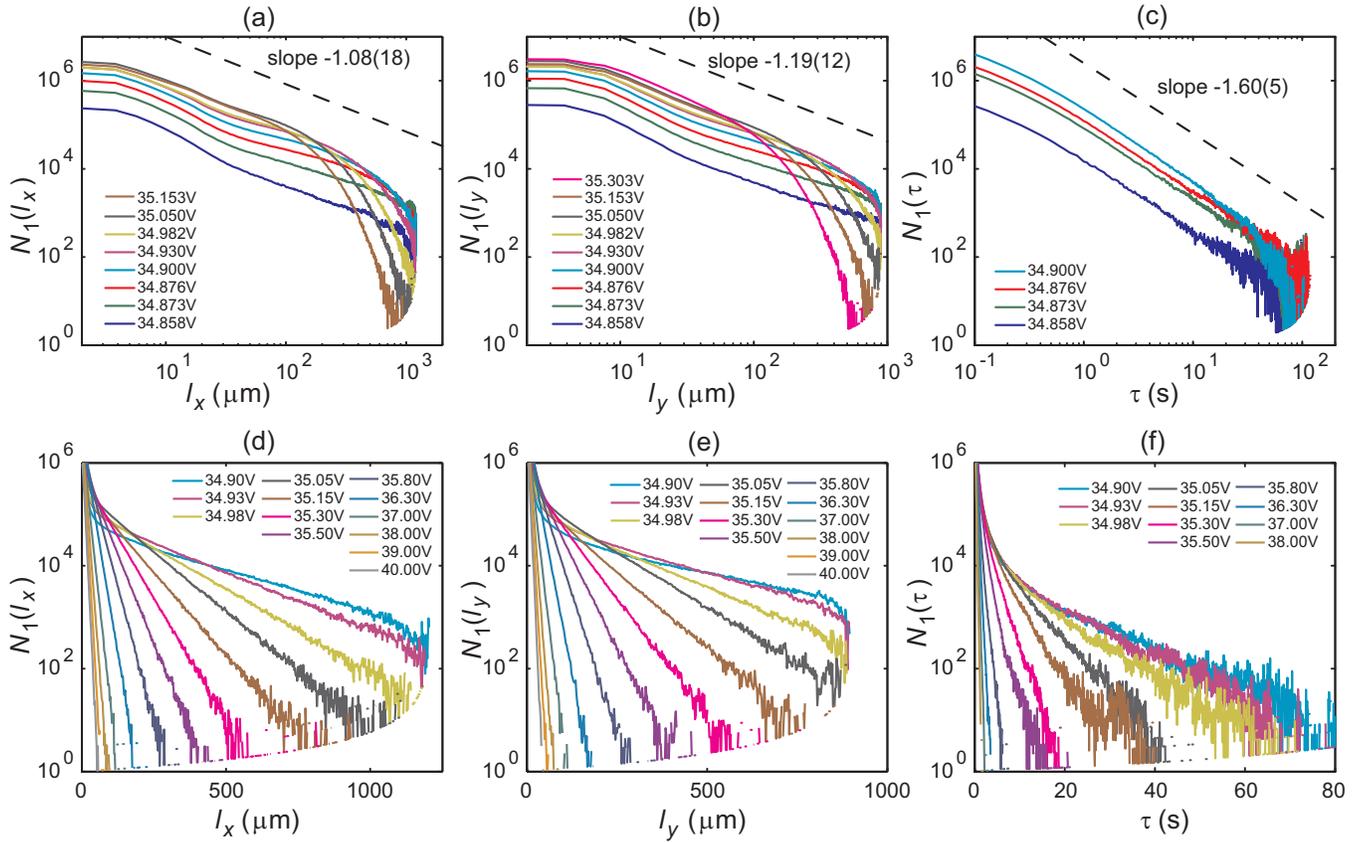}
 \caption{(Color online) Histograms of (inactive) DSM1 lengths $l_x, l_y$ and duration $\tau$ in the steady state, in double-logarithmic scales (a-c) and in semi-logarithmic scales (d-f). The applied voltages are in ascending order from bottom right to top left (a-c) and from top right to bottom left (d-f), respectively. Dashed lines show the estimated algebraic decay at criticality.}
 \label{fig:SteadyStateDSM1Dist}
\end{figure*}%

Figure \ref{fig:SteadyStateRho} shows
 the time-averaged order parameter $\bar{\rho}$,
 i.e., the average fraction of DSM2.
It shows that the transition is continuous
 and that $\bar{\rho}$ scales algebraically near the critical point.
Fitting these data with the usual scaling form \cite{Note1}
\begin{equation}
 \bar\rho \sim (V^2-V_{\rm c}^2)^{\beta},  \label{eq:DefinitionBeta}
\end{equation}
 the critical voltage $V_{\rm c}$ and the critical exponent $\beta$
 are found to be
\begin{equation}
 V_{\rm c} = 34.856 (4)\,\mathrm{V}, ~~~~\beta = 0.59(4),  \label{eq:SteadyStateRho}
\end{equation}
 where the numbers in parentheses indicate
 the range of errors in the last digits \cite{Note-Error}.
Our estimate $\beta = 0.59(4)$ is in good agreement
 with the value for $(2+1)$-dimensional DP,
 $\beta^\mathrm{DP} = 0.583(3)$
 \cite{Grassberger_Zhang-PhysA1996,Voigt_Ziff-PRE1997}.

We then measure the distributions of the sizes $l$ and durations $\tau$
 of inactive (DSM1) regions,
 or intervals between two neighboring active (DSM2) patches.
Histograms are made separately for each spatial direction, $l_x$ and $l_y$,
 to take account of the anisotropy of DSM1
 \cite{Strangi_etal-PRE1999,Nagaya_etal-JPSJ1999}.
Care is also taken to compensate missing intervals
 due to the finite observation window:
 since an interval of size $l$ may not be captured
 within a frame of size $L$,
 i.e., either edge of the interval may not be in the frame,
 with probability $l/L$,
 the unbiased distributions $N_1(l)$ are estimated from the observed ones
 by $N_1(l) = N_{\rm obs}(l) / (1-l/L)$ \cite{Note2}.

\begin{figure*}[t]
 \includegraphics[clip]{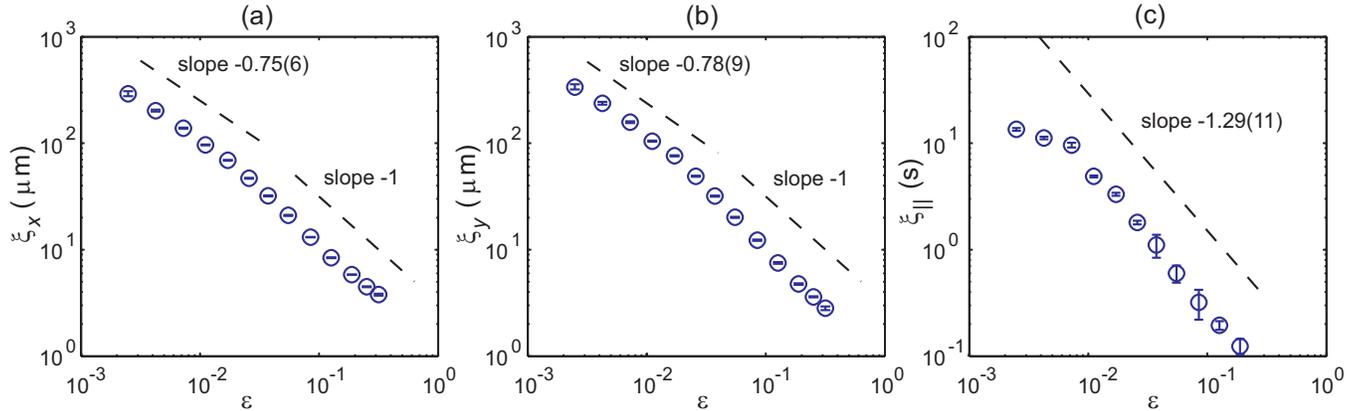}
 \caption{(Color online) Correlation length $\xi_x$, $\xi_y$ and correlation time $\xi_\parallel$ in the steady state, as functions of the deviation from criticality $\ep \equiv (V^2-V_{\rm c}^2)/V_{\rm c}^2$. Dashed lines are guides to eye.}
 \label{fig:SteadyStateNu}
\end{figure*}%

The results are shown in Fig.\ \ref{fig:SteadyStateDSM1Dist}.
We find that the DSM1 distributions decay algebraically
 within the observed length/time scale
 for voltages very close to criticality,
 while for higher voltages they start to decay exponentially
 from certain characteristic length/time scales,
 as expected for continuous absorbing phase transitions
 \cite{Henkel_etal-Book2008}.
The observed power-law decays
% which reflect spatial and temporal fractal properties of DSM2 clusters,
 are fitted as
 $N_1(l) \sim l^{-\mu_\perp}$ and $N_1(\tau) \sim \tau^{-\mu_\parallel}$
 with
\begin{equation}
 \mu_x = 1.08(18),~~~~\mu_y=1.19(12),~~~~\mu_\parallel=1.60(5),  \label{eq:SteadyStateMu}
\end{equation}
 where $\mu_x$ and $\mu_y$ indicate the exponent $\mu_\perp$
 measured in the $x$ and $y$ direction, respectively.
These exponents are directly connected to the fractal dimensions $d_{\rm f}$
 of the DSM2 clusters measured in the corresponding direction
 as $\mu = d_{\rm f} + 1$.
The critical exponents $\mu_\perp$ and $\mu_\parallel$ for DP
 can be estimated using scaling relations
 with the order parameter exponent $\beta$ and
 the correlation length/time exponents $\nu_\perp$ and $\nu_\parallel$,
 namely \cite{Henkel_etal-Book2008},
\begin{equation}
 \mu_\perp = 2-\beta/\nu_\perp,~~~~\mu_\parallel = 2-\beta/\nu_\parallel.  \label{eq:ScalingMu}
\end{equation}
They give $\mu_{\perp}^{\mathrm{DP}} = 1.204(2)$ and
 $\mu_{\parallel}^{\mathrm{DP}} = 1.5495(10)$ for $(2+1)$ dimensions
 \cite{Voigt_Ziff-PRE1997}.
Although the existence of slight discrepancies of order $10^{-2}$
 in Eq.\ \eqref{eq:ScalingMu} is suggested from numerical studies
 \cite{Hede_etal-JSP1991,Henkel_Peschanski-NuclPhysB1993},
 our estimates in Eq.\ \eqref{eq:SteadyStateMu}
 agree with values expected for the DP class at any rate.

Moreover, the DSM1 distributions allow
 estimating correlation length and time scales,
 $\xi_\perp$ and $\xi_\parallel$, respectively,
 from their exponential tails
 shown in Fig.\ \ref{fig:SteadyStateDSM1Dist}(d-f).
Fitting distributions with an empirical form
 \cite{Ciliberto_Bigazzi-PRL1988,Rupp_etal-PRE2003}
\begin{equation}
 N_1(l), N_1(\tau) \sim (Al^{-\mu} + B) \e^{-l/\xi},  \label{eq:DSM1DistEq}
\end{equation}
 with powers $\mu_\perp$ and $\mu_\parallel$ fixed at the estimates
 in Eq.\ \eqref{eq:SteadyStateMu},
 we obtain the results shown in Fig.\ \ref{fig:SteadyStateNu}.
Both $\xi_\perp$ and $\xi_\parallel$ show algebraic divergence
\begin{equation}
 \xi_\perp \sim (V^2 - V_{\rm c}^2)^{-\nu_\perp},~~~~\xi_\parallel \sim (V^2 - V_{\rm c}^2)^{-\nu_\parallel},  \label{eq:DefinitionNu}
\end{equation}
 near criticality as expected,
 except for the first two points in Fig.\ \ref{fig:SteadyStateNu}(c),
 which deviate from the power law
 presumably due to finite length of movies
 ($120\unit{s}$) used to count the distributions.
Fitting Eq.\ \eqref{eq:DefinitionNu} to the data in scaling regions
 (the first five points for $\xi_x,\xi_y$ and
 all the points except the first two for $\xi_\parallel$),
 we obtain
\begin{equation}
 \nu_x = 0.75(6),~~~~\nu_y = 0.78(9),~~~~\nu_\parallel = 1.29(11). \label{eq:SteadyStateNu}
\end{equation}
They are in good agreement with the DP values
 $\nu_{\perp}^{\mathrm{DP}}=0.733(3)$ and 
 $\nu_{\parallel}^{\mathrm{DP}}=1.295(6)$
 \cite{Grassberger_Zhang-PhysA1996,Voigt_Ziff-PRE1997}.
In addition, the facts that no significant anisotropy is found
 between $\xi_x$ and $\xi_y$ and
 that they can be even shorter than the cell depth $d=12\unit{\mu{}m}$
 [Fig.\ \ref{fig:SteadyStateNu}(a,b)]
 suggest that distributions of DSM2 patches are practically not influenced
 by the anisotropy of DSM1 \cite{Strangi_etal-PRE1999,Nagaya_etal-JPSJ1999}
 and by the existence of coherent roll structure
 of width roughly $d$ behind DSM1
 \cite{Nagaya_etal-JPSJ1999}.

\begin{figure*}[t]
 \includegraphics[clip]{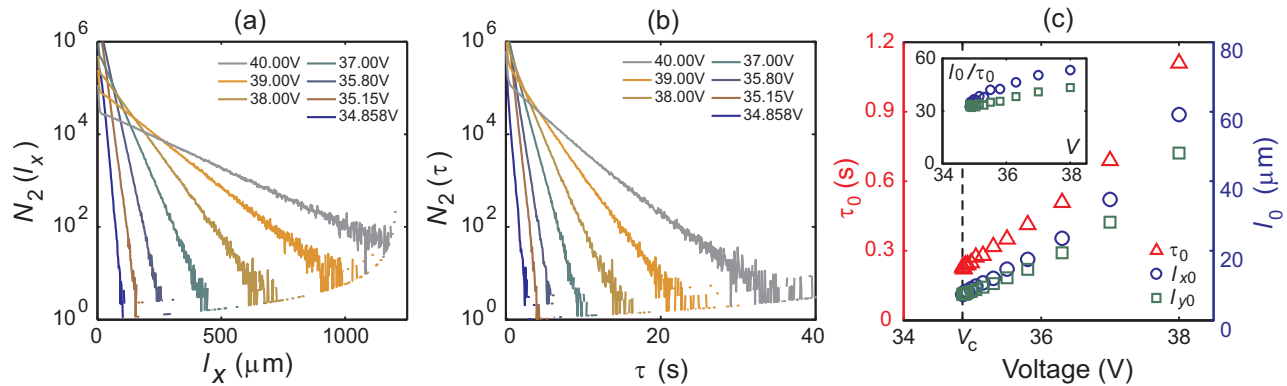}
 \caption{(Color online) Distributions of (active) DSM2 domain sizes in the steady state. (a,b) Histograms of DSM2 lengths $l_x$ in the $x$ direction (a) and of DSM2 duration $\tau$ (b). The applied voltages are in ascending order from bottom left to top right. (c) Characteristic length and time scales $l_{x0}, l_{y0}, \tau_0$, defined from the exponential decay in the histograms. The range of errors is smaller than the symbol size. Dashed line indicates the critical voltage $V_{\rm c}$. (Inset) Ratios of $l_{x0}$ and $l_{y0}$ to $\tau_0$ (same symbols as in the main panel).}
 \label{fig:SteadyStateDSM2Dist}
\end{figure*}%

On the other hand, distributions of sizes of active (DSM2) patches
 do not show any long range correlations even close to criticality
 [Fig.\ \ref{fig:SteadyStateDSM2Dist}(a,b)].
It implies that the local recession of DSM2 into DSM1
 sketched in Fig.\ \ref{fig:DSM1DSM2}(d) is indeed always present.
The effective relaxation rate can be directly estimated,
 just as an inverse of the characteristic time $\tau_0$
 from the exponential tail of the temporal DSM2 distribution,
 shown in Fig.\ \ref{fig:SteadyStateDSM2Dist}(c).
%This inverse rate
The characteristic time $\tau_0$
 increases linearly with $V$ except for high voltages,
 where it is not simply determined from the dynamics of individual patches
 because of the saturation of the DSM2 fraction (Fig.\ \ref{fig:SteadyStateRho}).
 In particular, it exhibits no sign of criticality.
All these observations are consistent with,
e.g., the dynamics of the so-called contact process \cite{Harris-AnnProb1974}
 (depicted in Fig.\ \ref{fig:DSM1DSM2}(d)),
a prototypical model showing a DP-class transition 
 \cite{Hinrichsen-AdvPhys2000,Henkel_etal-Book2008},
 indicating that such process indeed governs the 
 coarse-grained dynamics of our STI regime in liquid crystal turbulence.

Similarly, the spatial distributions of DSM2 sizes yield
 characteristic length scales $l_{x0}$ and $l_{y0}$
 [Fig.\ \ref{fig:SteadyStateDSM2Dist}(c)].
Although they appear to be also dominated by the relaxation process,
 as suggested from an effectively constant ratio $l_0/\tau_0$
 [inset of Fig.\ \ref{fig:SteadyStateDSM2Dist}(c)],
 they clearly show anisotropy [Fig.\ \ref{fig:SteadyStateDSM2Dist}(c)],
 as opposed to the correlation lengths
 $\xi_x$ and $\xi_y$ estimated from DSM1-size distributions
 [Fig.\ \ref{fig:SteadyStateNu}(a,b)].
This implies that the contamination process of DSM2 is indeed driven
 by the anisotropic, fluctuating shear flow of surrounding DSM1.
The larger effective contamination rate in $x$ is in line with the fact
 that the turbulent structure of DSM1 remains mainly in the $x$-$z$ plane
 \cite{Strangi_etal-PRE1999},
 but at odds with the global elliptic shape of growing DSM2 nuclei,
 longer in $y$,
 observed for higher voltages \cite{Kai_etal-JPSJ_PRL,Note3}.
Both would be explained if we assume that turbulent flow behind is
 faster and more correlated in the $y$ direction,
% as, indeed, roughly seen from movement of dust
% in badly-prepared cells,
 but further studies are necessary on this point.

\section{critical-quench experiment}  \label{sec:CriticalQuench}

A typical experiment performed usually on numerical models showing absorbing phase
transitions is the critical decay of active patches from fully active initial conditions
 \cite{Hinrichsen-AdvPhys2000,Henkel_etal-Book2008}.
In such critical-quench experiments,
 correlation length and time grow in time, and,
 as long as they remain much smaller than the system size,
 scaling estimates are free from finite-size effects.

Experiments are performed as follows:
 we first apply $60\unit{V}(\gg V_{\rm c})$ to the cell
 and wait until it is entirely invaded by DSM2 domains.
We then suddenly decrease the applied voltage to a value
 in the range of $34.86\unit{V} \leq V \leq 35.16\unit{V}$,
 i.e., near $V_{\rm c}$, and observe the time decay of activity
 for $900\unit{s}$.
We repeat this $10$ times for each $V$ and average the results
 over this ensemble.

% \subsection{Decay of order parameter}

\begin{figure}[t]
 \includegraphics[clip]{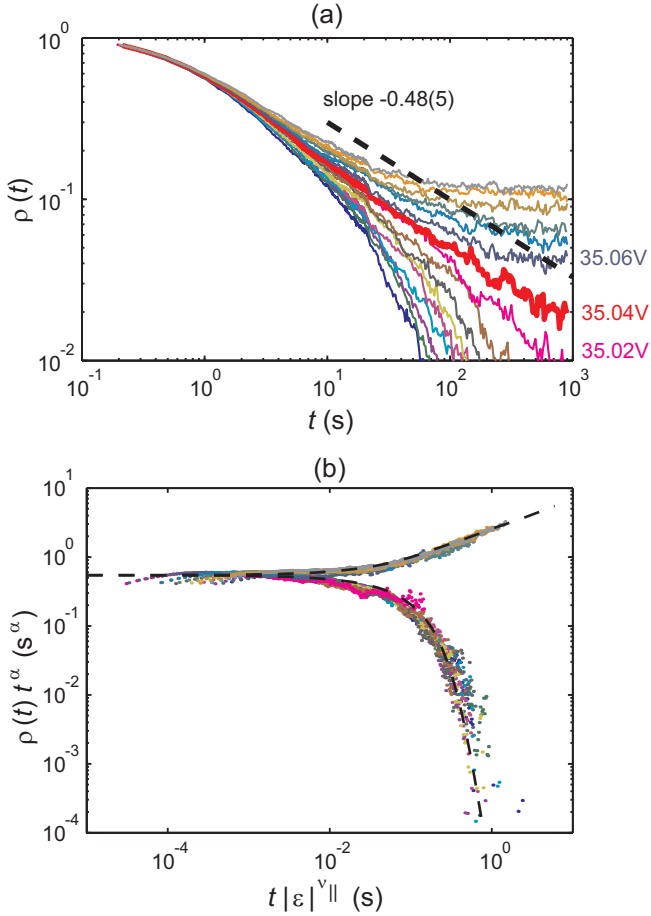}
 \caption{(Color online) Decay of the order parameter $\rho(t)$ on critical quenching. (a) $\rho(t)$ versus $t$, for $V=34.86\unit{V}$, $34.88\unit{V}$, $\cdots$, $35.16\unit{V}$ from bottom left to top right. The curve for $V=35.04\unit{V}$ (showing the longest power-law regime) is indicated by a thick line. (b) Same data with rescaled axes $t |\ep|^{\nu_\parallel}$ and $\rho(t)t^\alpha$, showing data collapsing. For $V_{\rm c}$, $\alpha$, and $\nu_\parallel$, we use values measured in the experiment [Eqs.\ \eqref{eq:SteadyStateNu} and \eqref{eq:CriticalQuenchAlpha}], but a collapse of similar quality is obtained also with DP-class exponent values. The dashed curve indicates the DP universal scaling function $F_{\rho}^\mathrm{DP}(x)$ obtained numerically from the contact process.}
 \label{fig:CriticalQuenchRho}
\end{figure}%

We first measure the decay of the order parameter $\rho(t)$ after the quench
 [Fig.\ \ref{fig:CriticalQuenchRho}(a)].
As expected, $\rho(t)$ decays exponentially with a certain characteristic time
 for lower voltages, converges to a finite value for higher voltages,
 and in-between at $V=35.04 \unit{V}$,
 it decays algebraically over the whole observation time.
A simple scaling ansatz implies the following functional form
 for $\rho(t)$ in this case:
\begin{equation}
 \rho(t) \sim t^{-\alpha} F_\rho(\ep t^{1/\nu_\parallel}),~~~~\alpha = \beta/\nu_\parallel,  \label{eq:ScalingFunctionRho}
\end{equation}
 where $\ep \equiv (V^2-V_{\rm c}^2)/V_{\rm c}^2$
 is the deviation from criticality
 and $F_\rho(x)$ is a universal scaling function.
From the slopes of the algebraic regimes for the three $V$ values
 closest to the threshold, we estimate
\begin{equation}
 V_{\rm c} = 35.04(1) \unit{V},~~~~\alpha = 0.48(5). \label{eq:CriticalQuenchAlpha}
\end{equation}
Note that $V_{\rm c}$ measured here is slightly higher than
 in the steady-state experiments.
In fact, the roll convection onset $V^* = 8.96\unit{V}$ was also higher.
We believe this is because of possible slight shift
 in the controlled temperature,
 and also of the ageing of our sample,
 a well known property of MBBA,
 during the days which separated the two sets of experiments.
On the other hand, no measurable shift of $V_{\rm c}$
 was detected during a given set of experiments.
We also confirmed that $V_{\rm c}$ here is consistent with
 a threshold roughly estimated from steady state
 just before the critical-quench experiments.
Our estimate of the critical exponent $\alpha$ is
 again in good agreement with the DP value
 $\alpha^\mathrm{DP} = 0.4505(10)$ \cite{Voigt_Ziff-PRE1997}.

Furthermore, the scaling form of Eq.\ \eqref{eq:ScalingFunctionRho}
 implies that the time series $\rho(t)$ for different voltages
 collapse on a single curve $F_\rho(x)$
 when  $\rho(t)t^\alpha$ is plotted as a function of 
 $t |\ep|^{\nu_\parallel}$.
Our data do collapse reasonably well [Fig.\ \ref{fig:CriticalQuenchRho}(b)],
 where the upper and lower branches
 correspond to $V>V_{\rm c}$ and $V<V_{\rm c}$, respectively.
It is compared and found in good agreement
 with the universal scaling function $F_\rho(x)$ of the DP class
 [dashed curve in Fig.\ \ref{fig:CriticalQuenchRho}(b)],
 calculated numerically from the $(2+1)$-dimensional contact process.
It shows that the decay of DSM2 patches is governed by this DP universal function,
 except for the very early stage
 where it is influenced by microscopic features of the liquid crystal, as expected.

\begin{figure}[t]
 \includegraphics[clip]{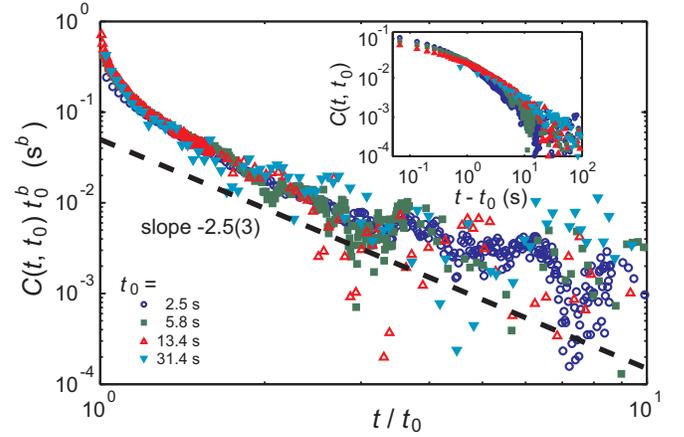}
 \caption{(Color online) Autocorrelation function $C(t,t_0)$ at $V = V_{\rm c}$, measured in the critical-quench experiments. Raw data in the inset are collapsed with rescaled axes $t/t_0$ and $C(t,t_0) t_0^b$. Dashed line shows the estimated asymptotic algebraic decay.}
 \label{fig:CriticalQuenchAgeing}
\end{figure}%

We also measure the autocorrelation function $C(t,t_0)$, defined as
\begin{equation}
 C(t,t_0) = \expct{\rho(\pr,t) \rho(\pr,t_0)} - \expct{\rho(\pr,t_0)} \expct{\rho(\pr,t)},  \label{eq:DefinitionCorrelator}
\end{equation}
 where $\expct{\cdots}$ denotes average in space and over ensembles.
During the critical decaying process,
 $C(t,t_0)$ is not a function of $t-t_0$
 (inset of Fig.\ \ref{fig:CriticalQuenchAgeing})
 but scaled rather by $t/t_0$,
 a feature sometimes referred to as ``ageing'' in the literature
 \cite{Henkel-JPhysCM2007}.
Our data can be collapsed (Fig.\ \ref{fig:CriticalQuenchAgeing})
 using the expected scaling form
\begin{equation}
 C(t,t_0) \sim t_0^{-b} F_C (t/t_0), ~~F_C(x) \sim x^{-\lambda_C/z} ~~(x \to \infty),  \label{eq:ScalingCorrelator}
\end{equation}
 with
\begin{equation}
 b = 0.9(1), ~~~~ \lambda_C/z = 2.5(3).  \label{eq:CriticalQuenchAgeingExp}
\end{equation}
Both are in agreement with DP values
% $b^\mathrm{DP} = 0.9$ and $\lambda_C^\mathrm{DP}/z^\mathrm{DP} = 2.8(3)$
% found numerically \cite{Ramasco_etal-JPhysA2004}, and with
 $b^\mathrm{DP} = 0.901(2)$ and
 $\lambda_C^\mathrm{DP}/z^\mathrm{DP} = 2.583(14)$,
 estimated from the scaling relations \cite{Baumann_Gambassi-JStatMech2007}
\begin{equation}
 b = 2\beta/\nu_{\parallel},  ~~~~\lambda_C/z = 1 + (\beta + d_{\rm s}\nu_\perp) / \nu_{\parallel},  \label{eq:ScalingBLambdaC}
\end{equation}
 where $d_{\rm s} = 2$ is the spatial dimension.

% \subsection{Local persistence}

The dynamic aspect of critical behavior can also be characterized
using first-passage quantities
 \cite{Majumdar-CurrSci1999}.
One of these is the local persistence probability
 $P_{\rm l}(t)$, defined as the probability that
the local state at a given point in space has not changed
 until time $t$,
which typically shows a power-law decay 
$P_{\rm l}(t) \sim t^{-\theta_{\rm l}}$.
The non-trivial local persistence exponent $\theta_{\rm l}$
is known to be in general independent of usual critical exponents
 such as $\beta, \nu_\perp, \nu_\parallel$,
 and far less is known about its universality,
 mainly due to the fact that persistence is a quantity
 involving an infinite-point correlation function.

In the context of DP,
 local persistence is measured from the probability that
 initially inactive ``sites'' do not become active up to time $t$. 
(The persistence of activity is always dominated
 by the local relaxation into the inactive state
 and shows only exponential decay,
 similarly to Fig.\ \ref{fig:SteadyStateDSM2Dist}.)
Initial conditions are typically set to be random in numerical studies
 \cite{Hinrichsen_Koduvely-EPJB1998,Albano_Munoz-PRE2001,Menon_etal-EPL2003,Fuchs_etal-JStatMech2008},
 which is, however, impossible in this experiment at present.
We therefore consider, as ``initial condition'' for the local persistence,
configurations chosen at some moment $t_0$ during the critical quench
 and measure the persistence probability $P_{\rm l}(t-t_0)$.

\begin{figure}[t]
 \includegraphics[clip]{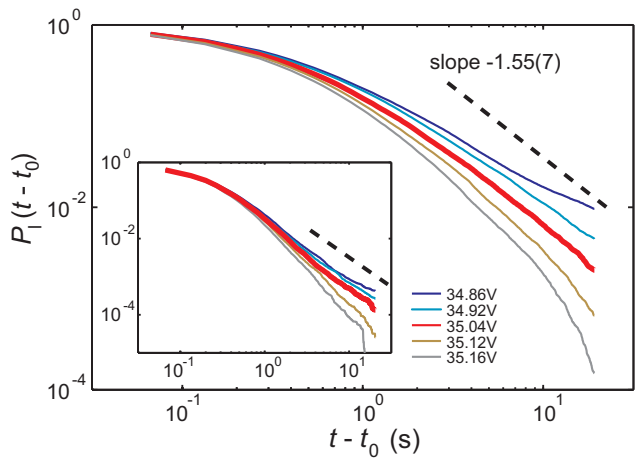}
 \caption{(Color online) Local persistence probability $P_{\rm l}(t)$ in the critical-quench experiments. The initial time is $t_0 = 2.14 \unit{s}$ for the main panel and $t_0 = 0.81 \unit{s}$ for the inset. The applied voltages are in ascending order from top to bottom. Dashed lines are guides to eye with the same slope.}
 \label{fig:CriticalQuenchPersistence}
\end{figure}%

The scaling regime for the order parameter decay observed in 
Fig.~\ref{fig:CriticalQuenchRho} starting only after times of typically
$1$-$2 \unit{s}$, we further limit ourselves to $t_0$ values larger than this 
microscopic, non-universal time.
Typical results are shown for $t_0 = 2.14 \unit{s}$
 in Fig.\ \ref{fig:CriticalQuenchPersistence}.
The local persistence probability $P_{\rm l}(t)$ is found to
 converge to a finite constant for $V < V_{\rm c}$
 and to decay exponentially for $V > V_{\rm c}$,
 as expected from numerical studies of absorbing phase transitions
 \cite{Hinrichsen_Koduvely-EPJB1998,Albano_Munoz-PRE2001,Menon_etal-EPL2003,Fuchs_etal-JStatMech2008,Grassberger-arXiv2009}.
At criticality, $P_{\rm l}(t)$ decays algebraically (again beyond some time
of the order of $t_0$) with 
\begin{equation}
 \theta_{\rm l} = 1.55(7).  \label{eq:CriticalQuenchThetal}
\end{equation}
The value of $\theta_{\rm l}$
 for $(2+1)$-dimensional DP is still a matter of debate,
 with different estimates in past numerical studies:
 $\theta_{\rm l}^\mathrm{DP} = 1.50(1)$ \cite{Albano_Munoz-PRE2001},
 $\theta_{\rm l}^\mathrm{DP} \gtrsim 1.6$ \cite{Fuchs_etal-JStatMech2008},
 and very recently $\theta_{\rm l}^\mathrm{DP} = 1.611(7)$,
 found with an improved algorithm \cite{Grassberger-arXiv2009}.
%This agrees
% with estimates from numerical studies,
% $\theta_{\rm l}^\mathrm{DP} = 1.50(1)$ \cite{Albano_Munoz-PRE2001},
% and also with $\theta_{\rm l}^\mathrm{DP} = 1.611(7)$,
% \cite{Grassberger-arXiv2009}.
%(This value is still debated, though, since
% Ref.\ \cite{Fuchs_etal-JStatMech2008} was led to only provide
% lower bounds $\theta_{\rm l}^\mathrm{DP} > 1.62$ 
% and $\theta_{\rm l}^\mathrm{DP} > 1.58$
% for $(2+1)$-dimensional models.)
Our value, which is rather robust with respect to changing
 the initial conditions and the initial time
 (Fig.\ \ref{fig:CriticalQuenchPersistence}), is in agreement
 with all numerical estimates to our accuracy.

\section{critical-spreading experiment}  \label{sec:CriticalSpreading}

In order to complete the characterization of the dynamic critical behavior
 of the DSM1-DSM2 transition, we performed
critical spreading experiments,
 which start from a single seed of active, DSM2 region.
This allows to measure other critical exponents, such as the one
 governing the scaling of the probability $P_{\rm s}(\infty)$
 that a cluster starting from a single active seed
 survives forever: $P_{\rm s}(\infty) \sim \ep^{\beta'}$,
 known to serve as another order parameter
 characterizing absorbing phase transitions
 \cite{Hinrichsen-AdvPhys2000,Henkel_etal-Book2008}.

The two exponents $\beta$ and $\beta'$ are
% in general different, but
 known to be equal to each other for the DP class,
 thanks to an extra symmetry linked to time-reversal,
 the so-called rapidity symmetry
 \cite{Hinrichsen-AdvPhys2000,Henkel_etal-Book2008}.
This symmetry also implies that $\alpha = \delta$
 where $\delta$ is defined from the time decay of the survival 
 probability at threshold, $P_{\rm s}(t) \sim t^{-\delta}$.

We stress, however, that the rapidity symmetry
 and the resulting scaling relations 
 do not hold generically in absorbing phase transitions.
Although
% critical scaling of hysteresis loops in the DSM1-DSM2 transition
% suggests a value of $\beta'$ consistent with DP \cite{Takeuchi-PRE2008},
 a value of $\beta'$ consistent with DP was suggested
 from the critical scaling of hysteresis loops in the DSM1-DSM2 transition
 \cite{Takeuchi-PRE2008},
 it remains important to assess all spreading exponents as accurately and
independently as possible, in order to complete our characterization
of DP-class critical behavior and check directly the rapidity symmetry.

\subsection{Technique for nucleating DSM2}

In contrast to numerical simulations,
 it is not easy, in experiments,
to prepare an initial, single, localized seed of DSM2
 in an otherwise homogeneous DSM1 system.
We developed an experimental technique for nucleating
 a DSM2 patch artificially, using a pulse laser.

\begin{figure}[t]
 \includegraphics[clip]{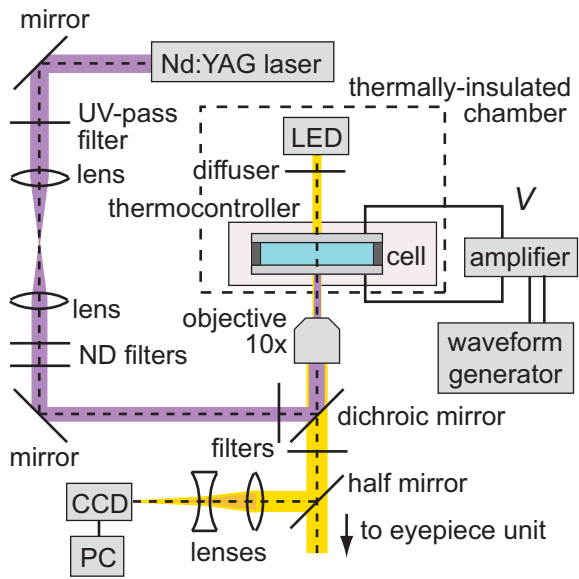}
 \caption{(Color online) Schematic diagram of the experimental setup for the critical-spreading experiment. UV: ultraviolet, ND: neutral density. See text for details.}
 \label{fig:ExpSetupSpreading}
\end{figure}%

\begin{figure}[t]
 \includegraphics[clip]{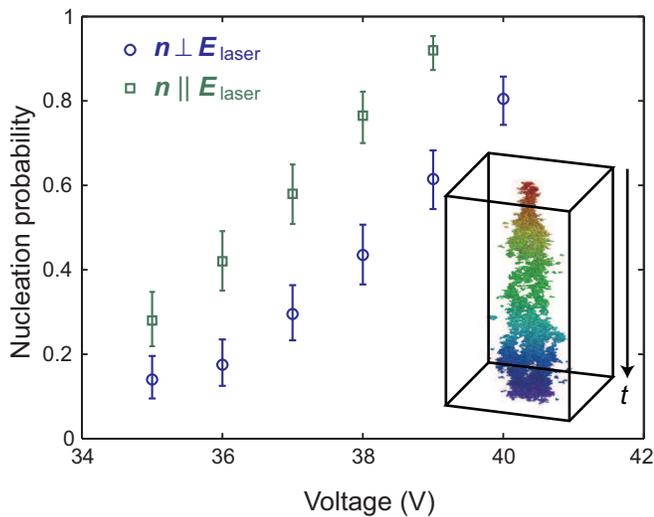}
 \caption{(Color online) Probability of DSM2 nucleation induced by five successive pulses of laser. Laser is linearly polarized along or perpendicularly to the mean direction $\bm{n}$ of the molecules (squares and circles, respectively). Error bars denote the $95\%$ confidence intervals assuming the binomial distribution. (Inset) Typical spatiotemporal diagram of nucleated DSM2 cluster for $V = 36.65\unit{V} > V_{\rm c}$. Scales are $500\unit{\mu{}m} \times 500\unit{\mu{}m}$ in space and $30\unit{s}$ in time.}
 \label{fig:DSM2Nucleation}
\end{figure}%

The experimental setup for the critical-spreading experiments is
 schematically shown in Fig.\ \ref{fig:ExpSetupSpreading}.
We emit $4$-$6\unit{ns}$ pulses of Nd:YAG laser
 (MiniLase II 20Hz, New Wave Research),
 focused by an objective lens
 ($\times 10$, NA $0.30$, UPlanFLN, Olympus),
 into the cell.
Using its third harmonic at $355\unit{nm}$,
 around which MBBA has a broad absorption band
 \cite{Mizuno_Shinoda-MolCrystLiqCryst1978, Mizuno_Shinoda-MolCrystLiqCryst1981},
 and reducing its energy to roughly
 $0.3 \unit{nJ} \approx 2 \times 10^9 \unit{eV}$ at the cell,
 we can indeed nucleate DSM2 from the absorbing DSM1 state
 (inset of Fig.\ \ref{fig:DSM2Nucleation})
 without any observable damage to the sample.
Figure \ref{fig:DSM2Nucleation} shows the probability
 of the DSM2 nucleation induced by emitting five successive laser pulses
 at $20\unit{Hz}$.
The nucleation probability increases with voltage
 and is significantly higher for laser polarized along the mean director field.
This confirms that nucleation is indeed brought about
 by the laser absorption of MBBA,
 since its absorbance is higher along the long axis of the molecule
 \cite{Mizuno_Shinoda-MolCrystLiqCryst1978, Mizuno_Shinoda-MolCrystLiqCryst1981}.
Moreover, the electronic structure of MBBA reveals that
 the ultraviolet absorption band stems from the C--N and C--C bonds
 between the aniline and benzyliden rings
 and is strongly coupled with twist angles there
 \cite{Mizuno_Shinoda-MolCrystLiqCryst1981}.
We therefore speculate that absorption of ultraviolet laser pulses
 might lead to a sudden conformation change in the molecular structure,
 creating a topological defect and thus triggering a DSM2 nucleation.

\subsection{Results}

\begin{figure*}[t]
 \includegraphics[clip]{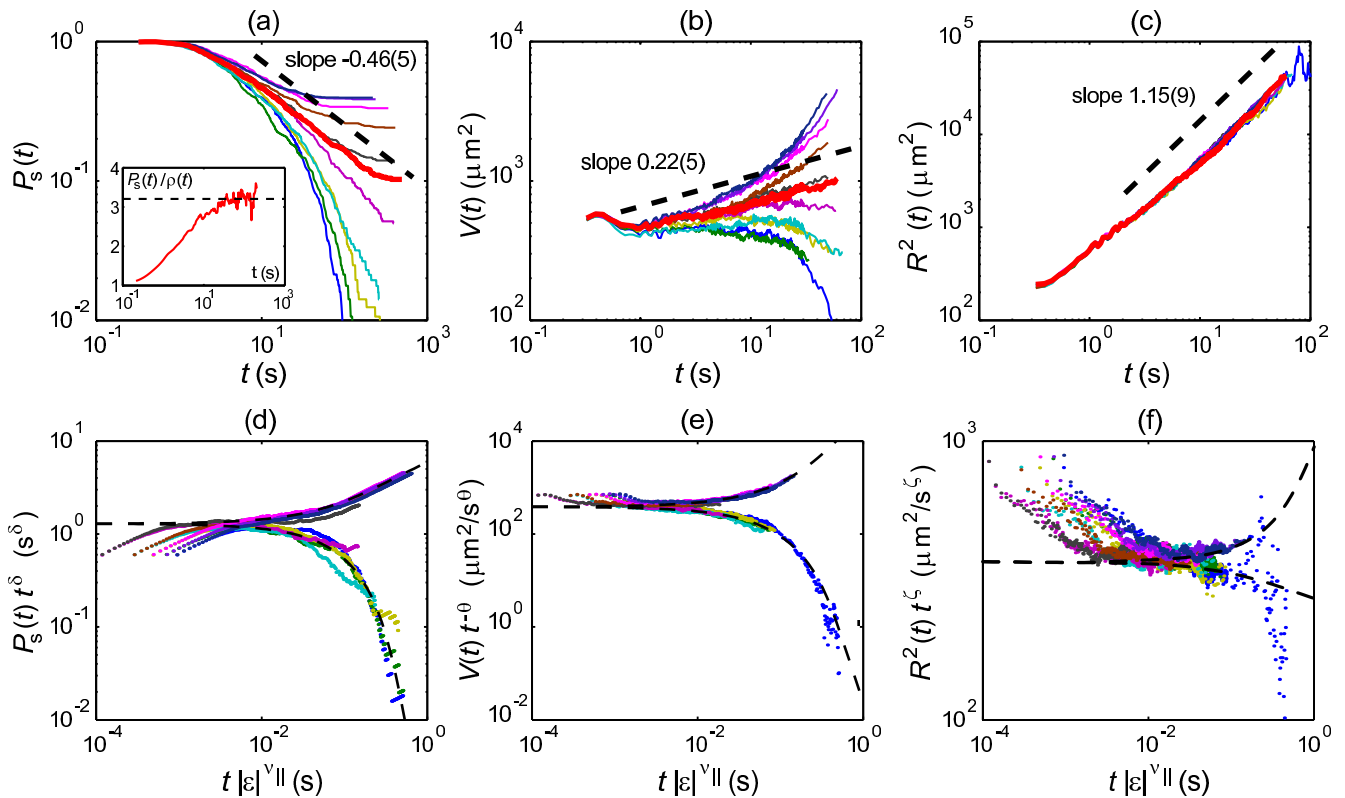}
 \caption{(Color online) Results of the critical-spreading experiments. (a-c) Survival probability $P_{\rm s}(t)$, volume $V(t)$, and mean square radius $R^2(t)$ of clusters started from a single DSM2 nucleus, for $V = 36.25\unit{V}, 36.29\unit{V}, \cdots, 36.65\unit{V}$ from bottom to top. The data for $R^2(t)$ are mostly overlapping. The curves for $V = 36.45\unit{V} = V_{\rm c}$ are drawn with thick lines. [Inset of (a)] The ratio of the survival probability $P_{\rm s}(t)$ from the critical-spreading experiments to the DSM2 fraction $\rho(t)$ from the critical-quench experiments, both at criticality. The dashed line is a guide to eye, indicating roughly the asymptotic ratio of $P_{\rm s}(t) / \rho(t)$. (d-f) Same data with axes scaled after the expected scalings \eqref{eq:ScalingFunctionP}-\eqref{eq:ScalingFunctionR}. We use values of $V_{\rm c}$ and critical exponents measured in the experiments. The dashed curves indicate the DP universal scaling functions $F_{\rm s}^\mathrm{DP}(x), F_{\rm v}^\mathrm{DP}(x), F_{\rm r}^\mathrm{DP}(x)$ obtained numerically from the contact process.}
 \label{fig:CriticalSpreadingResults}
\end{figure*}%

We perform the critical-spreading experiments with the technique above.
For each measurement we emit ten successive pulses
 polarized along the molecules (in the $x$ direction),
 with which DSM2 is always nucleated for voltages of interest.
In order to make reliable statistics for the survival probability,
 it is necessary to repeat experiments at least hundreds of times
 for each voltage.
This led us to improve further the temperature control of the cell,
 both in short and long time scales.
This is achieved by placing the thermocontroller
 in a thermally-insulated chamber
 (Fig.\ \ref{fig:ExpSetupSpreading}),
 made of plastic foam stage, wall, and ceiling,
 whose temperature inside is kept constant
 by circulating constant-temperature water.
The cell temperature is finally stabilized at $25.0\unit{^\circ{}C}$
 with fluctuations at most $2\unit{mK}$ over two weeks.
This allows us to repeat the experiment $563$-$567$ times for each voltage
 in the range of $36.25\unit{V} \leq V \leq 36.65\unit{V}$.
The roll convection onset and
 the critical voltage roughly measured in steady state
 were $V^* = 8.55\unit{V}$
 and $V_{\rm c} \approx 36.5\unit{V}$, respectively.

We measure not only the survival probability $P_{\rm s}(t)$
 but also the volume $V(t)$ and the mean square radius $R^2(t)$
 of DSM2 clusters, averaged over all the repetitions,
 even if the cluster dies before time $t$.
The following relations are then expected from the scaling ansatz
 \cite{Hinrichsen-AdvPhys2000,Henkel_etal-Book2008}:
\begin{align}
 &P_{\rm s}(t) \sim t^{-\delta} F_{\rm s}(\ep t^{1/\nu_\parallel}),  &&\delta = \beta' / \nu_\parallel,  \label{eq:ScalingFunctionP} \\
 &V(t) \sim t^{\theta} F_{\rm v}(\ep t^{1/\nu_\parallel}),  &&\theta = (d_{\rm s} \nu_\perp - \beta - \beta') / \nu_\parallel,  \label{eq:ScalingFunctionV} \\
 &R^2(t) \sim t^\zeta F_{\rm r}(\ep t^{1/\nu_\parallel}),  &&\zeta = 2/z = 2\nu_\perp / \nu_\parallel,  \label{eq:ScalingFunctionR}
\end{align}
 where $F_{\rm s}(x), F_{\rm v}(x), F_{\rm r}(x)$
 are universal scaling functions.

The experimental results are shown in Fig.\ \ref{fig:CriticalSpreadingResults}.
Except for $R^2(t)$ where all the data are almost overlapping,
 which is also typically the case in simulations of DP-class models,
 the data show opposite curvatures below and above a certain voltage.
Seeking for the curve with the longest algebraic regime in $V(t)$,
 which is statistically most reliable,
 we locate the critical voltage $V_{\rm c}$ at
\begin{equation}
 V_{\rm c} = 36.45(2) \unit{V}.  \label{eq:CriticalSpreadingVc}
\end{equation}
We then measure the three critical exponents $\delta,\theta,\zeta$
 from the algebraic regime for three voltages around $V_{\rm c}$, yielding
\begin{equation}
 \delta = 0.46(5), ~~~~\theta = 0.22(5), ~~~~\zeta = 1.15(9).  \label{eq:CriticalSpreadingExp}
\end{equation}
All of them are in good agreement with the DP exponents,
 $\delta^\mathrm{DP} = 0.4505(10), \theta^\mathrm{DP} = 0.2295(10)$,
 and $\zeta^\mathrm{DP} = 1.1325(10)$ \cite{Voigt_Ziff-PRE1997}.
This confirms the scaling relations expressing the rapidity symmetry. 
It is further confirmed by plotting
the ratio of the survival probability $P_{\rm s}(t)$
 to the DSM2 fraction $\rho(t)$ in the critical-quench experiments
 [inset of Fig.\ \ref{fig:CriticalSpreadingResults}(a)], which shows that
 the two order parameters become asymptotically proportional to each other
\begin{equation}
 P_{\rm s}(t) \sim m^2 \rho(t)  \label{eq:RapiditySymmetry2}
\end{equation}
 with the coefficient $m^2 \approx 3.2$.

We also tested data collapse [Fig.\ \ref{fig:CriticalSpreadingResults}(d-f)],
 which provides reasonable results given the
 limited statistical accuracy of the data.
The obtained scaling functions coincide satisfactorily with those
of the $(2+1)$-dimensional contact process
 [dashed curves in Fig.\ \ref{fig:CriticalSpreadingResults}(d-f)].
As in the data collapse of $\rho(t)$
 [Fig.\ \ref{fig:CriticalQuenchRho}(b)],
 the collapsed data show typical time scales
 above which the corresponding quantities are governed
 by the DP scaling functions.

\section{summary and discussion}  \label{sec:Summary}

We have performed three series of experiments, namely,
 steady-state experiments, critical-quench experiments,
 and critical-spreading experiments,
 to characterize the critical behavior of the DSM1-DSM2 transition
 in liquid crystal turbulence.
Table \ref{tbl:SummaryExponents} summarizes the main results.
We have measured in total 12 critical exponents
 with reasonable accuracy,
 typically over a few orders of magnitude of power-law regimes.
All of the measured exponent values agree within a few percent
 with those defining the DP universality class.
Given that most of them are theoretically linked through scaling relations,
 we can equivalently say that we have experimentally confirmed
 those scaling relations, 8 in total, that connect the measured exponents
 (Table \ref{tbl:SummaryScalingRelations}).
Among them, we have confirmed in particular the rapidity symmetry
 $\alpha = \delta$, providing also the asymptotic amplitude
 of the ratio between the two order parameters, $m^2 \approx 3.2$.
Moreover, we have also tested the expected scaling forms
 of Eqs.\ \eqref{eq:ScalingFunctionRho}, \eqref{eq:ScalingCorrelator},
 \eqref{eq:ScalingFunctionP}-\eqref{eq:ScalingFunctionR}
 through data collapse, and found them in good agreement with
 numerically-obtained DP universal scaling functions.
Based on all these results, we conclude that the DSM1-DSM2 transition
 constitutes an unambiguous experimental realization
 of an absorbing phase transition in the DP universality class.

\begin{table}[t]
 \caption{Summary of the measured critical exponents.}
 \label{tbl:SummaryExponents}
 \catcode`?=\active \def?{\phantom{0}}
 \begin{tabular}{lllll} \hline \hline
  \multicolumn{2}{l}{Exponent} & \multicolumn{2}{l}{DSM1-DSM2\footnotemark[1]} & DP\cite{Grassberger_Zhang-PhysA1996,Voigt_Ziff-PRE1997} \\ \hline
  density order parameter & $\beta$ & $0.59(4)$ & & $0.583(3)$ \\
  correlation length & $\nu_\perp$ & $0.75(6)$ & $0.78(9)$ & $0.733(3)$ \\
  correlation time & $\nu_\parallel$ & $1.29(11)$ & & $1.295(6)$ \\
  inactive interval in space & $\mu_\perp$ & $1.08(18)$ & $1.19(12)$ & $1.204(2)$\footnotemark[2] \\
  inactive interval in time & $\mu_\parallel$ & $1.60(5)$ & & $1.5495(10)$\footnotemark[2] \\
  density decay & $\alpha$ & $0.48(5)$ & & $0.4505(10)$ \\
  local persistence & $\theta_{\rm l}$ & $1.55(7)$ & & $1.611(7)$ \cite{Grassberger-arXiv2009} \\
  ageing in autocorrelator & $b$ & $0.9(1)$ & & $0.901(2)$ \\
  & $\lambda_C / z$ & $2.5(3)$ & & $2.583(14)$ \\
  survival probability & $\delta$ & $0.46(5)$ & & $0.4505(10)$ \\
  cluster volume & $\theta$ & $0.22(5)$ & & $0.2295(10)$ \\
  cluster mean sqr. radius & $\zeta$ & $1.15(9)$ & & $1.1325(10)$ \\ \hline \hline
 \end{tabular}
 \footnotetext[1]{For $\mu_\perp$ and $\nu_\perp$ exponents measured in $x$ and $y$ direction are shown in this order.}
 \footnotetext[2]{See also the remark \cite{Note4}.}
\end{table}

\begin{table}[t]
 \caption{Experimentally confirmed scaling relations.}
 \label{tbl:SummaryScalingRelations}
 \catcode`?=\active \def?{\phantom{0}}
 \begin{tabular}{lllll} \hline \hline
  \multicolumn{2}{l}{Scaling relations} & LHS & RHS & DP\cite{Grassberger_Zhang-PhysA1996,Voigt_Ziff-PRE1997}\\ \hline
  $\mu_\perp = 2 - \beta/\nu_\perp$\footnotemark[1] & (in $x$) & $1.08(18)$ & $1.21(8)$ & $1.204(2)$ \\
  & (in $y$) & $1.19(12)$ & $1.24(10)$ & $1.204(2)$ \\
  $\mu_\parallel = 2 - \beta/\nu_\parallel$\footnotemark[1] & & $1.60(5)$ & $1.54(5)$ & $1.5495(10)$ \\
  $\alpha = \beta/\nu_\parallel$ & & $0.48(5)$ & $0.46(5)$ & $0.4505(10)$ \\
  $b = 2\beta/\nu_\parallel$ & & $0.9(1)$ & $0.91(10)$ & $0.901(2)$ \\
  \multicolumn{2}{l}{$\lambda_C / z = 1 + (\beta + d_{\rm s}\nu_\perp) / \nu_{\parallel}$} & $2.5(3)$ & $2.62(17)$\footnotemark[2] & $2.583(14)$ \\
  $\delta = \beta/\nu_\parallel$ & & $0.46(5)$ & $0.46(5)$ & $0.4505(10)$ \\
  $\theta = (d_{\rm s} \nu_\perp - 2\beta) / \nu_\parallel$ & & $0.22(5)$ & $0.25(11)$\footnotemark[2] & $0.2295(10)$ \\
  $\zeta = 2/z = 2\nu_\perp / \nu_\parallel$ & & $1.15(9)$ & $1.16(14)$\footnotemark[2] & $1.1325(10)$ \\ \hline\hline
 \end{tabular}
 \footnotetext[1]{See also the remark \cite{Note4}.}
 \footnotetext[2]{The value of $\nu_x$ is used for $\mu_\perp$.}
\end{table}

We now return to our initial remark concerning the 
surprising scarcity of experimental realizations
of DP-class transitions (Table \ref{tbl:EarlierExperiments}).

One central difficulty lies in the necessity to avoid, as much as possible,
quenched disorder, which is known to be relevant.
It is theoretically known that such disorder
 does affect DP criticality and even destroys it for strong disorder
 \cite{Hooyberghs_etal-PRL2003}.
Recent theoretical and numerical studies show that
 even weak disorder changes the asymptotic critical behavior
 \cite{Hoyos-PRE2008,Vojta_etal-PRE2009,Fallert_Taraskin-PRE2009},
 but the characteristic length/time scale separating
 DP and disordered, so-called ``activated'' critical behavior
 grows fast with decreasing strength of disorder
 \cite{Hoyos-PRE2008,Vojta_etal-PRE2009}.
It is therefore important to work with systems made of macroscopic units,
 where quenched disorder is expected to be negligible.
% Most, if not all previous experiments do so.
We consider that the quenched disorder in our system,
which may take the form of inhomogeneities in the electrodes or 
impurities in the sample, is also sufficiently weak.

Still, our DSM1-DSM2 transition seems to be the only fully convincing one.
Three factors explain, in our view, why our experiments provided such
clear DP scaling laws.

(a) \textit{Large system size and fast response}.
As already mentioned, one great advantage to work with electroconvection
 is that very large aspect ratios can easily be realized.
The number of effective degrees of freedom of our cell is $2.7 \times 10^6$,
 which is orders of magnitude larger than any earlier experiment
 (Table \ref{tbl:EarlierExperiments}).
This considerably suppresses finite-size effects and allows to observe
scaling on several orders of magnitude.
Similarly, the typical microscopic timescales of liquid crystals 
are very short (of the order of $10\unit{ms}$),
 providing accurate statistics in reasonable laboratory time.

(b) \textit{Almost perfectly absorbing state}.
The condition for being an absorbing state,
 i.e., that the system can \textit{never} escape once it entered,
 appears to be somewhat too strict from the experimental point of view.
Indeed, spontaneous nucleation of the active state
 seems to have been present at least in some of the past experiments
 \cite{Hinrichsen-AdvPhys2000,Rupp_etal-PRE2003}, which blurs the critical
behavior beyond some finite scales.

On the other hand, our active state, DSM2, consists of topological defects,
 whose spontaneous formation is in principle forbidden.
Of course this may occur in practice,
 as suggested from the observation of vanishing hysteresis
 at the DSM1-DSM2 transition \cite{Takeuchi-PRE2008},
 but the nucleation rate remains so low that we cannot directly observe it,
 constituting an almost perfectly absorbing state.
%%%XXX is this true? do we have spontaneous  nucleation in the bulk with steady
%% control parameters? (i.e. not during a ramp like for hysteresis scaling)
%% if not, then our AS is really ``perfect''

(c) \textit{Fluctuating absorbing state}.
In most earlier experiments and numerical studies,
 the absorbing state has been essentially fluctuation-free, or laminar.
This is indeed a natural choice suggested from the minimal theory of DP,
 and causes of course no problem in numerical studies.
In experiments, however, such absorbing states may typically cause
 long-range effects through the rigidity of their laminar pattern
 and/or the propagation of soliton-like objects,
 which may further reduce the effective system size
 and even break DP scaling
 \cite{Hinrichsen-AdvPhys2000,Henkel_etal-Book2008}.
In contrast, our absorbing state, DSM1,
 is itself a fluctuating, turbulent state.
Long-range interactions are then likely to be killed
 by the local turbulent fluctuations of DSM1,
 which may have led to the observation of clean DP critical behavior.
It is noteworthy to remark that the Chat\'e-Manneville coupled map lattice
 \cite{Chate_Manneville-PhysD1988},
 a deterministic numerical model for absorbing phase transitions
 with a non-chaotic (laminar) absorbing state,
 exhibits non-DP critical behavior
 probably due to soliton-like objects propagating through laminar regions.
Interestingly, an elementary modification to render its absorbing state
 itself chaotic does lead to DP scaling \cite{Chate-unpublished}.

In conclusion, we have experimentally found that the DSM1-DSM2 transition
 in the electroconvection of nematic liquid crystals,
 showing an absorbing transition into spatiotemporal intermittency,
 clearly belongs to the DP universality class.
Both static and dynamic critical behavior has been investigated
 with the help of the newly developed experimental technique
 to create a seed of DSM2,
 confirming a total of 12 critical exponents, 5 scaling functions,
 8 scaling relations, and in particular the rapidity symmetry,
 in full agreement
 with those characterizing the DP class in $2+1$ dimensions.
We hope that this first clear and comprehensive experimental realization
 of a DP-class transition
will trigger further studies of absorbing phase transitions and of related
situations with genuinely non-equilibrium  critical properties.
In this respect the recent works of Cort\'e {\it et al.}
 \cite{Corte_etal-NatPhys2008}
 and of Mangan {\it et al.} \cite{Mangan_etal-PRL2008}
 are especially encouraging,
 concerning experiments and realistic situations,
 respectively,
 for absorbing transitions with a conserved field.

\acknowledgments

The authors are grateful to I. Dornic, F. Ginelli, J. A. Hoyos,
 S. Kai, and N. Oikawa for fruitful discussions.
We would also like to thank M. Henkel and G. I. Menon
 for drawing our attention
 to ageing and Ref.\ \cite{Henkel_Peschanski-NuclPhysB1993}
 and to local persistence, respectively.
This work is supported in part
 by Grant-in-Aid for Scientific Research (18068005) and for JSPS Fellows.

\appendix*
\section{Binarizing images}

Every analysis presented in the paper is performed
 using binarized images, where DSM2 domains are distinguished
 from the absorbing DSM1 background.
We give here a detailed description of the binarizing algorithm we used.

The binarization is carried out in the following manner:
(a) We prepare three successive images taken at 15 frames per second,
 and remove the inhomogeneity of the incident light intensity.
(b) We then normalize the obtained intensity $I$ of the three images
 with respect to mean $\langle I_\mathrm{DSM1} \rangle$
 and standard deviation $\delta I_\mathrm{DSM1}$ of the DSM1 intensity
 at a given voltage, namely
$I_\mathrm{norm} = (I-\langle I_\mathrm{DSM1} \rangle)/\delta I_\mathrm{DSM1}$.
Note that we can separately measure the intensity of the fully DSM1 state
 even above the threshold $V_{\rm c}$,
 since DSM1 always appears first when the voltage is applied.
(c) Since DSM2 domains have lower transmittance than DSM1,
 we extract the regions where the normalized intensity
 is less than a certain threshold $-I_\mathrm{th}$.
Here we chose $I_\mathrm{th} = 1.5$,
 determined so as to obtain a good agreement with
 direct visual observations, particularly in movies.
(d) Taking into consideration that DSM2 domains move much slower
 than the local intensity fluctuations in DSM1
 (recall that DSM1 is itself a turbulent state)
 and that the minimum DSM2 area is $d^2/2$ \cite{Kai_etal-JPSJ_PRL},
 where $d$ is the depth of the cell,
 we take the logical intersection (``AND'' operator)
 of the three successive images,
 and then remove clusters with area smaller than $d^2/2$.
Clusters are screened out in this way,
 and their binarized images in the middle frame are used as final images.
In other words, the intersection is used
 only for comparison with the minimum area.
(e) Finally, we cut off the periphery of the image of width $d/2\sqrt{2}$, 
 since this region is biased in the step (d).
The size of the binarized images
 reduces to $1206 \unit{\mu{}m} \times 899 \unit{\mu{}m}$,
 which roughly corresponds to $142 \times 106$ effective degrees of freedom.
We confirmed that the chosen threshold ($I_\mathrm{th} = 1.5$)
 works well all over the range of voltages we investigate,
 and that no DSM2 region is falsely detected
 when binarizing images of the fully DSM1 phase.
Typical results of the binarization are shown
 in Fig.\ \ref{fig:DSM1DSM2}(c) and (e), and Movie S2 of Ref.\ \cite{EPAPS},
 where we can confirm that DSM2 domains are precisely detected.

For the critical-spreading experiments
 presented in Sec.\ \ref{sec:CriticalSpreading},
 we have slightly modified the binarization algorithm
 in order to detect DSM2 patches
 originating only from the prepared seed,
 and not to miss them.
To this end, we have reduced the intensity threshold $I_\mathrm{th}$ to $1.2$,
 and instead, binarized only within a target region,
 defined from positions of DSM2 patches in previous images
 (position of the seed for the first image)
 and assumed maximal displacement of DSM2,
 which is chosen to be much larger than the actual displacement,
 namely, $d$ during successive two images taken at $15$ fps
 and extrapolated diffusively.
We again confirmed that binarized images from a single set of parameters
 closely follow the actual evolution of DSM2 patches
 for all the voltages of interest.
A typical result is shown in the inset of Fig.\ \ref{fig:DSM2Nucleation}.

\end{document}